\documentstyle[12pt,aasms4]{article}
\def\ltsima{$\; \buildrel < \over \sim \;$}
\def\simlt{\lower.5ex\hbox{\ltsima}}
\def\gtsima{$\; \buildrel > \over \sim \;$}
\def\simgt{\lower.5ex\hbox{\gtsima}}
\def\s{\ifmmode \widetilde \else \~\fi}
\def\={\overline}

\def\spose#1{\hbox to 0pt{#1\hss}}

\def\etal{{\it et al.\ }}
\def\cf{{\it cf.\ }}
\def\eg{{ e.g.,\ }}

\def\lta{\mathrel{\spose{\lower 3pt\hbox{$\mathchar"218$}}
     \raise 2.0pt\hbox{$\mathchar"13C$}}}
\def\gta{\mathrel{\spose{\lower 3pt\hbox{$\mathchar"218$}}
     \raise 2.0pt\hbox{$\mathchar"13E$}}}
\def\Dt{\spose{\raise 1.5ex\hbox{\hskip3pt$\mathchar"201$}}}	
\def\dt{\spose{\raise 1.0ex\hbox{\hskip2pt$\mathchar"201$}}}	

\def\=={\equiv}

\def\dotsfill{\leaders\hbox to 1em{\hss.\hss}\hfill}

\slugcomment{Submitted to The Astronomical Journal}

\lefthead{Ibata \& Irwin}
\righthead{Discrete Classification with PCA}

\begin{document}

\title{Discrete Classification with Principal Component Analysis: \nl
Discrimination of Giant and Dwarf Spectra in K-stars}

\author{Rodrigo A. Ibata}
\affil{Department of Physics and Astronomy, University of British Columbia, \nl
2219 Main Mall, Vancouver, B.C., V6T 1Z4, Canada \nl
Electronic mail: ibata@astro.ubc.ca}
 
\author{Michael J. Irwin}
\affil{Royal Greenwich Observatory, Madingley Rd, Cambridge CB3 0EZ \nl
Electronic mail: mike@ast.cam.ac.uk}



\begin{abstract}
We demonstrate  the use of a variant  of  Principal Component Analysis (PCA)
for discrimination  problems in astronomy.  This  variant of PCA is shown to
provide the   best linear discrimination  between  data classes.   As a test
case,  we present the problem  of discrimination between K~giant and K~dwarf
stars from  intermediate resolution  spectra near  the  Mg `b'  feature. The
discrimination procedure is trained on a set of 24  standard K~giants and 24
standard K~dwarfs, and then used to perform giant -- dwarf classification on
a sample of  $\approx 1500$ field  K~stars of unknown luminosity class which
were   initially classified visually.   For  the  highest S/N spectra,   the
automated classification agrees very well (at the 90 -- 95\% level) with the
visual classification. Most importantly, however,   the automated method  is
found to classify stars in a repeatable fashion, and, according to numerical
experiments, is very robust to signal to noise (S/N) degradation.

\end{abstract}




%

\section{Introduction}

Studies  of the large  scale kinematic and chemical  structure of our Galaxy
often investigate  the properties of samples  of stars  that are believed to
trace Galactic mass.  One  of  the most  used  species is  the K~giant  (\eg
\markcite{Ib95a} Ibata   \&  Gilmore  1995a,  \markcite{Rich90}   Rich 1990,
\markcite{Rich88} Rich   1988,  \markcite{Kui89}  Kuijken \&   Gilmore 1989,
\markcite{Lew89} Lewis   \&  Freeman 1989).   These  stars  are particularly
useful as  they  give a   fair  representation of   the underlying   stellar
distribution (\cf e.g. \markcite{Ib95a} Ibata \&  Gilmore 1995a).  They have
high intrinsic     luminosity and  well-behaved    spectral features   whose
differences can be interpreted in terms of changes in temperature, abundance
and  surface gravity.   However,  local,  intrinsically faint  K~dwarfs  can
contaminate  these  samples considerably.   Fortunately,  the sensitivity of
absorption  lines  to  the  surface gravity of  these   stars allows  one to
discriminate between K~giants and  K~dwarfs: this may be performed visually,
by  comparison to  a grid  of   standards (\eg  \markcite{Kui89} Kuijken  \&
Gilmore  1989),  or by minimizing  a  statistic constructed from the stellar
spectra and a   grid  of synthetic standards  (\eg  \markcite{Cay91a} Cayrel
\etal 1991a, \markcite{Cay91b} 1991b).

A number  of alternative methods based on  real stellar templates  have been
used for spectral classification including:  Artificial Neural Networks (\eg
\markcite{Hip94}   von Hippel  \etal  1994);  minimum  distance  methods and
assorted  methods   based on  cross-correlation  (\eg \markcite{Kur84} Kurtz
1984).   Neural   networks   can offer   a  very   sophisticated  non-linear
combination  of input parameters  and can   be thought of   as a  variant of
non-linear least   squares   minimization closely    tied    to a   Bayesian
classification scheme.    While  cross-correlation and the  closely  related
(weighted)  minimum  distance  methods   are  straightforward variants    of
least-squares fitting to standard templates. In this note we demonstrate the
use of a robust optimal  linear discrimination scheme based  on a variant of
Principal Components Analysis.

The  data-set that   is examined   below was   obtained  with  the   AUTOFIB
multi-fibers spectrograph at  the Anglo Australian  Telescope (AAT) with the
aim of investigating the kinematic and abundance structure of both the inner
Milky Way (\markcite{me95a} Ibata \& Gilmore 1995a, \markcite{me95b} 1995b),
and the Sagittarius dwarf galaxy  (\markcite{IGI94} Ibata, Gilmore \&  Irwin
1994).  Though this  instrument was efficient in  gathering large samples of
spectroscopic data, the   resulting spectra cannot  be directly  compared to
flux-calibrated    spectra,   because  the    spectrograph   induces  large,
low-frequency variations in the shape  of the spectra over wavelength ranges
of  width  typically $\approx  200 \AA$.   This  meant that  the giant-dwarf
discrimination technique of \markcite{Cay91a}  Carrel \etal (1991) could not
be applied to the AUTOFIB data without considerable recalibration.

Ibata \& Gilmore (1995a) therefore initially classified their $\approx 3000$
spectra  visually, following the  prescription  detailed in \markcite{Kui89}
Kuijken \& Gilmore (1989).   However, it was clearly  desirable to design an
automated algorithm that  is repeatable,  that   classifies stars to   lower
signal to  noise than is possible visually,  and that allows  an estimate of
the certainty of the classification to be made.

\section{Visual Dwarf -- Giant Classification for K stars}

The survey-star spectra were first compared empirically to a grid of K giant
and    K  dwarf  standards.    The   standard  spectra   were   observed  by
\markcite{Kui89} Kuijken \&  Gilmore (1989)  and \markcite{me95a}  Ibata  \&
Gilmore (1995a), again with the AUTOFIB fibers system.  Standards at several
${\rm (B-V)_0}$ are presented in Figure~1; a list of these stars is given in
Table~1.    The  field dwarfs  and  metal  poor  dwarfs (subdwarfs) are from
\markcite{Bes79} Bessel  \&  Wickramasinghe  (1979)   and   \markcite{Rod74}
Rodgers \& Eggen   (1974), the metal  rich dwarfs  (Hyades dwarfs)  are from
\markcite{Pels75}  Pels \etal (1975)   and \markcite{Upg77} Upgren \&  Weiss
(1977),  while  the giants  are  from  \markcite{Yoss81} Yoss  \etal (1981),
\markcite{Friel86} Friel  (1986) and  \markcite{Fab85} Faber  \etal  (1985).
The most  striking features in  these spectra are  the three Mg`b'  lines at
(5167,  5173 and 5184 $\AA$)  and  the MgH  band at  5211  $\AA$ (which also
belongs  to the Mg`b' feature).   The other prominent  lines are mostly TiO,
\hbox{Fe I} and \hbox{Fe II}.  Several properties of  K star atmospheres can
be seen in the grid.  In dwarfs, the prominent MgH band (5211 $\AA$) is seen
after ${\rm  (B-V)_0 \simgt 1.05}$, while in   giants it appears  only after
${\rm  (B-V)_0 \simgt 1.25}$.  Fe lines  are weaker even in super-metal-rich
giants (\cf Table~1) than  in Hyades  dwarfs of  the same color.   For those
stars with ${\rm (B-V)_0 > 1.1}$, the wide Mg`b' absorption band (a wide dip
stretching ${\rm 5050 \AA \simlt \lambda \simlt 5200}$) is strong in dwarfs,
but is weak until ${\rm (B-V)_0 \approx 1.3}$ in giants.

\markcite{Cay91} Cayrel \etal (1991) calculate synthetic spectra to find the
surface gravity dependence of   a K star spectrum   in the wavelength  range
$4800$ to $5300 \AA$  at fixed  effective temperature  and  metallicity.  In
this situation  they show that  dwarfs display much stronger  Mg, Fe and MgH
lines than giants (because giants have lower surface gravity atmospheres and
hence lower opacities).      Cayrel \etal also  calculate  the   metallicity
dependence at constant  surface  gravity and  effective  temperature  --- as
would be expected,  higher metallicity increases the depth  of the Mg and Fe
lines   and the MgH  band (except   for saturated  Mg  lines  in  metal poor
dwarfs). Their results show clearly that the Mg`b' triplet  and MgH band are
more sensitive  to gravity  than to  metallicity  for stars of ${\rm  [Fe/H]
\simgt -1.25}$, and that these lines can be as weak  in metal poor dwarfs as
they are in giants.   Fortunately, along the lines  of sight to these survey
stars, starcount galaxy models  predict a negligible contribution  ($<$ 0.01
\%  )    of metal poor    dwarfs  (foreground halo   stars)   in the samples
(\markcite{me94} Ibata 1994).

K giants and K dwarfs were in this  way visually classified by comparison to
the standards in Figure~1. The spectra were  also binned into four groups, a
subjective ranking of the  certainty of the classification.  The giant-dwarf
classification was deemed  to be satisfactory for  high S/N spectra, but was
clearly  unsatisfactory on noisy  spectra (judging from repeated attempts at
classification), especially on the  bluer end of  the selection range (${\rm
(B-V)_0 \simlt 1.0}$).

\section{Technique}

Below, we first remind  the reader of the  standard PCA technique,  and then
describe    the variant of   this  method  which  was  successfully  used to
discriminate between K~giants and K~dwarfs.

To begin with, the spectra to be analyzed are shifted into their rest frames
and binned linearly over  a fixed wavelength  range (4800-5500 $\AA$) into a
fixed number of bins $N$ (500).  Each spectrum can  thus be represented as a
point in the $N$-dimensional vector space of all possible (similarly binned)
spectra.

As an example,  consider the set of $n_d$  standard dwarf spectra.  This set
can be represented as a cloud of $n_d$ points in the above vector space. The
aim  of  the  standard PCA  classification    scheme is  to  concentrate the
information in the $n_d$   $N$-dimensional points into  a  set of $q$ ($q  <
n_d$)   orthogonal  $N$-dimensional  vectors   which are   able  to describe
``dwarf-ness'' to good approximation (in a least squares sense). The largest
of these vectors ${\bf a}_1$ is the direction along which the cloud of dwarf
stars is most elongated, that is, the direction of a  least squares line fit
to the dwarf points that passes though  the mean point (mean spectrum). This
is the first order  least squares description of  the data. The variation of
spectra in the direction ${\bf a}_1$ is the greatest in the  data set, so it
is removed by collapsing the cloud of points along ${\bf a}_1$ to give a new
data-set of dimension   $N-1$. The second  order   least squares description
${\bf  a}_2$ is calculated from  this new data set in  the same way as ${\bf
a}_1$ was from the original set. This process  is iterated so that the $i$th
principal  component is calculated from   a data-set formed by  successively
collapsing the  original data-set along  the 1st to the  ($i-1$)th principal
component directions. The maximum number $q_{max}$ of  such vectors that can
be  found  is either  $N$  (there are only $N$    dimensions to collapse the
data-set  into)  or   $n_d$  (when   all   points lie  exactly   along   the
($q_{max}=n_d$)th principal component).  If dwarf star spectra  have regular
patterns, the cloud  of points in the vector  space will be localized, so we
expect to be able to account for most of  the variance in  the sample with a
small number of principal components and the  aim of the operation will have
been fulfilled.

It can  easily  be shown  (\eg  \markcite{Fran91}  Francis  1991) that  this
process  is  equivalent to   finding the  eigenvectors corresponding  to the
largest eigenvalues of the matrix
\begin{equation}
{\bf \rm C} = \sum_i ( {\bf x}_{k} ) ({\bf x}_{k})^{\rm H},
\end{equation}
where ${\rm H}$ denotes  Hermitian conjugate, and  ${\bf x}_k$ is the  $k$th
sample vector.

The  problem that needs to   be addressed however,  is  how to  discriminate
between  classes of spectra (or  clouds  of points  in  the vector space  of
possible spectra).  The variant of PCA employed  here does not deconstruct a
single set of  spectra (cloud of points) as  above, but instead deconstructs
the set of difference vectors between points of  different classes (again in
a least squares sense) (see \eg Ullman 1973 \markcite{Ullman-1973}). We will
denote ${\bf x}_k^{(\mu)}$ as the $k$th sample  vector of class $\mu$. Since
the  mean  spectrum    ${\bf \overline{x}}$   contains  no    discriminatory
information, we  first subtract   ${\bf \overline{x}}$  from all  the  ${\bf
x}_k^{(\mu)}$: this does not affect discrimination and avoids the problem of
the mean spectrum dominating the covariance matrix (Equation~10 below) which
can make the eigenvector equation (Equation~9)  unsuitable for solution with
simple numerical algorithms.

Define a linear transformation $A$, such that
\begin{equation}
{\bf y}_k^{(\mu)} = A^{\rm H} {\bf x}_k^{(\mu)},
\end{equation}
where ${\bf y}_k^{(\mu)}$ is to be set up such that  it contains the maximum
amount  of   discriminatory  information   based on    a   least-mean-square
representation of all  the difference vectors  between the sets.   Let ${\bf
a}_i$ be the vector elements of the transformation matrix
$A=({\bf a}_1,\ldots,{\bf a}_i,\ldots,{\bf a}_M)$,
where $M$  is a fixed   number of elements less  than  or equal  to $N$.  We
therefore seek unit vectors ${\bf a}_i$ that maximize the quantity
\begin{equation}
S^2 = \sum_{k,l,\mu,\mu' \atop \mu \ne \mu'} \Bigl( {\bf y}_k^{(\mu)} - 
{\bf y}_l^{(\mu')}\Bigr)^2.
\end{equation}
The requirement that the ${\bf a}_i$ be unit vectors imposes the constraint
$({\bf a}_i)^{\rm H} {\bf a}_i = 1$.
Then, using  Lagrange multipliers, we  may  write:
\begin{equation}
{{d}\over{d{\bf a}_i}} 
\Biggl[
\sum_{k,l,\mu,\mu' \atop \mu \ne \mu'}
\Bigl( {\bf y}_k^{(\mu)} - {\bf y}_l^{(\mu')}\Bigr)^{\rm H} 
\Bigl( {\bf y}_k^{(\mu)} - {\bf y}_l^{(\mu')}\Bigr) - 
\lambda_i {\bf a}_i^{\rm H} {\bf a}_i 
\Biggr] = 0,
\end{equation}
where  $\lambda_i$ is the  Lagrange multiplier for ${\bf a}_i$. Substituting
for ${\bf y}_k^{(\mu)}$ from Equation~2:
\begin{equation}
{{d}\over{d{\bf a}_i}} 
\Biggl[
\sum_{k,l,\mu,\mu' \atop \mu \ne \mu'} 
\Bigl({\bf x}_k^{(\mu)} - {\bf x}_l^{(\mu')}\Bigr)^{\rm H} 
\Bigl({\rm A} {\rm A}^{\rm H} \Bigr) 
\Bigl({\bf x}_k^{(\mu)} - {\bf x}_l^{(\mu')}\Bigr) - 
\lambda_i {\bf a}_i^{\rm H} {\bf a}_i \Biggr] = 0.
\end{equation}
Putting ${\bf b}_{kl}^{(\mu \mu')}=( {\bf x}_k^{(\mu)} - {\bf x}_l^{(\mu')})$:
\begin{equation}
{{d}\over{d{\bf a}_i}} 
\Biggl[\sum_{k,l,\mu,\mu' \atop \mu \ne \mu'}
\Bigl({\bf b}_{kl}^{(\mu \mu')}\Bigr)^{\rm H} {\rm A} {\rm A}^{\rm H} ~ 
{\bf b}_{kl}^{(\mu\mu')} - \lambda_i {\bf a}_i^{\rm H} {\bf a}_i 
\Biggr] = 0.
\end{equation}
\begin{equation}
\sum_{k,l,\mu,\mu' \atop \mu \ne \mu'}
{{d}\over{d{\bf a}_i}}
\Biggl[
\Bigl( 
{\bf a}_1 \cdot {\bf b}_{kl}^{(\mu \mu')},
\ldots,
{\bf a}_i \cdot {\bf b}_{kl}^{(\mu \mu')},
\ldots,
{\bf a}_n \cdot {\bf b}_{kl}^{(\mu \mu')}
\Bigr)
\pmatrix{
{\bf a}_1 \cdot {\bf b}_{kl}^{(\mu \mu')}\cr
\vdots\cr
{\bf a}_i \cdot {\bf b}_{kl}^{(\mu \mu')}\cr
\vdots\cr
{\bf a}_n \cdot {\bf b}_{kl}^{(\mu \mu')}\cr
}
- \lambda_i {\bf a}_i^{\rm H} {\bf a}_i 
\Biggr] = 0.
\end{equation}
Differentiating:
\begin{equation}
2 \Biggl[\sum_{k,l,\mu,\mu' \atop \mu \ne \mu'} 
{\bf b}_{kl}^{(\mu \mu')} ~
\Bigl({\bf b}_{kl}^{(\mu \mu')}\Bigr)^{\rm H} - 
\lambda_i {\rm I} \Biggr] {\bf a}_i = 0,
\end{equation}
or:
\begin{equation}
[{\bf \rm C} - \lambda_i {\rm I}] {\bf a}_i = 0.
\end{equation}
Therefore the ${\bf a}_i$ are eigenvectors of the Hermitian matrix 
${\bf \rm C}$:
\begin{equation}
{\bf \rm C} = \sum_{k,l,\mu,\mu' \atop \mu \ne \mu'} 
\Bigl( {\bf x}_k^{(\mu)} - {\bf x}_l^{(\mu')} \Bigr) 
\Bigl( {\bf x}_k^{(\mu)} - {\bf x}_l^{(\mu')} \Bigr)^{\rm H},
\end{equation}
which is simply the covariance matrix of the difference vectors. The
eigenvectors 
${\bf a}_1,\ldots,{\bf a}_i,\ldots,{\bf a}_M$ of the covariance matrix define
the linear transformation $A$ (Equation~2 above).

Substituting Equation~9 into Equation~3 and using the orthogonality property
of the eigenvectors, we find
\begin{equation}
S^2 = \sum_{i-1}^m \lambda_i,
\end{equation}
where  $m$   is the  no. of  eigenvectors    used.  Clearly the  larger  the
eigenvector  the   larger  the discriminating   power  of  the corresponding
eigenvector. Therefore, the $M$  eigenvectors  with the largest  eigenvalues
give the  best $M$-dimensional  linear discrimination between  classes $\mu$
and $\mu'$.

\section{Results}

We   first find the  mean  spectrum of the   48  standard stars displayed in
Figure~1.  The covariance matrix ${\bf \rm C}$ (Equation~10) is formed using
the mean-corrected standard star  spectra.  The eigenvectors ${\bf a}_i$ and
related eigenvalues $\lambda_i$  of the covariance matrix  ${\bf \rm C}$ are
then calculated.  As explained  above, the eigenvectors corresponding to the
largest   eigenvalues  are those which    contain   the greatest amount   of
discriminatory  information:  the  ten  largest  eigenvalues  are  given  in
Table~2.   From the table we see  that the first order discriminating vector
(that  with the  largest  eigenvalue)  accounts  for a   large fraction  ---
$\approx  50$\% ---   of  the  total   discrimination, with  the   next nine
contributing only a further $\approx 35$\%. We therefore investigate whether
the first  order eigenvector  is  sufficient to allow  distinction between K
dwarfs and   K  giants.  This eigenvector   is shown  in  Figure~2:  several
spectral features of K stars are visible, notably the Mg`b' feature, the MgH
band at $5205\AA$ and some prominent Fe lines.

For each  star spectrum ${\bf x}$  that  is to  be classified, a coefficient
$c={\bf (x-{\overline s}) \cdot e}$ is calculated from the eigenvector ${\bf
e}$   (shown in  Figure~2)   and the average  standard  star  spectrum ${\bf
\overline s}$.  Figure~3 shows the  relation between ${\rm (B-V)_0}$ and $c$
calculated for the standard stars; the `stars'  in the diagram represent the
giants in the sample of standards, the large  dots represent dwarf stars and
the  small dots represent subdwarfs.  Superimposed  is a straight line drawn
by eye marking the boundary between giants and dwarfs.   For ${\rm (B-V)_0 >
1.1}$,  the  classification scheme appears   to work well.  For  ${\rm (B-V)
\simlt 1.0}$, the classification is  less  clear cut (but   it is also  very
difficult to classify these hotter stars visually --- \cf Section~2).

Figure~4 shows the same plot for the survey stars, where the ${\rm (B-V)_0}$
values  have  been taken from the  calibrated  APM photometry  (\cf Ibata \&
Gilmore 1995a).

We now estimate  the accuracy of  this classification scheme.  The effect of
photon   noise on classification is investigated   by degrading the standard
star spectra using a Poisson  random number generator  by Press \etal (1986)
\markcite{Press-etal-1986}. We find an $rms$ error in the coefficient $c$ of
$\approx 10$ when the  signal to noise is  degraded to ${\rm S/N \approx 5}$
(for the standard stars, $c$  takes values $ -100 \simlt  c \simlt 100 $). A
very  much  larger source of  error in   the classification of  survey stars
arises simply from the $rms$ color error ${\rm (B-V) \approx 0.18}$ of these
data (Ibata \& Gilmore 1995a \markcite{me95a}). Assuming  that each point in
Figure~4  has a probability  density that  is  a Gaussian distribution  with
$\sigma=0.18$ along the ${\rm (B-V)}$ direction, we find that $\approx 15$\%
of giants and $\approx 25$\% of dwarfs are on average misidentified.

Comparing the results of automated classification to that performed visually
(\cf   Section~2), we  find  that  $\approx  11$\% of all  dwarfs classified
visually are classified differently by the PCA  algorithm ($\approx 5$\% for
high signal to noise spectra with ${\rm (B-V) > 1.1}$), while $\approx 13$\%
of giants  classified visually are classified   differently by the algorithm
($\approx 8$\%  for high signal to noise  spectra with ${\rm (B-V) > 1.1}$).
(By high  signal to noise spectra we  mean approximately  the quarter of the
survey sample of which we were most confident of the visual classification).
We cannot easily  quantify the relative  precision between the automatic and
visual  techniques  since  it   is  non-trivial to   organize  a  controlled
experiment  on humans.   However,  in the  few   cases where  an inter-  and
intra-comparison between classification by   human experts and an  automated
algorithm has  been carried out for related  problems (\eg Naim \etal\ 1995,
Lahav \etal\ 1995 for galaxy morphology classification), the scatter between
different human experts  was found to be  non-neglible and indeed comparable
to the error from the automatic technique.  What is  clear from our tests is
that repeated human attempts at visual classification of low signal to noise
($S/N  \approx 10$) spectra are  much less reliable   than the machine based
approach (although again it is difficult to quantify this statement).

\section{Conclusions}

We discussed a variant of the  Principal Component Analysis technique, which
is designed  to  discriminate  between classes  of objects.  This  technique
provides the  best possible linear   discrimination.  It is  well suited  to
astronomical problems involving the  discrimination of spectra. We show that
it is  very   simple  to implement   this   technique  on   the problem   of
distinguishing   K~dwarfs   from K~giants  using    spectra  sampled  in the
wavelength range $4800$  to $5300 \AA$ at  $\approx 1.5 \AA$  resolution. In
principle, with very accurate ${\rm (B-V)_0}$ photometry, and with very high
signal to  noise  spectra   (say,   $S/N \simgt  30$)  it   is  possible  to
discriminate  visually   between     K~giants   and   K~dwarfs    to    high
accuracy. However, the K~star sample investigated  above had poor photometry
($\delta{\rm  (B-V)_0} \approx 0.2$), and  many  spectra had $S/N \simlt 10$
(for which repeated    attempts  at visual discrimination     gave different
results).  The  numerical algorithm  developed is  able  to reproduce visual
discrimination of the highest  signal to noise  spectra ($S/N \simgt 20$) to
approximately 90  --   95\% (and  helped  to  pick  out stars  which,   with
hindsight,  had   been  obviously  misclassified).   According to  numerical
experiments, in which standard star spectra had  their signal to noise ratio
degraded artificially to $S/N  \approx 5$ (about  the lowest $S/N$  spectrum
obtained),  the algorithm works well  with poor quality spectra. The machine
discrimination is reliable (the discrimination criteria remain fixed) and is
much more   accurate than  visual  discrimination on    low signal to  noise
spectra.

\clearpage

\begin{figure}
\epsfxsize=\hsize
\centerline{\epsffile{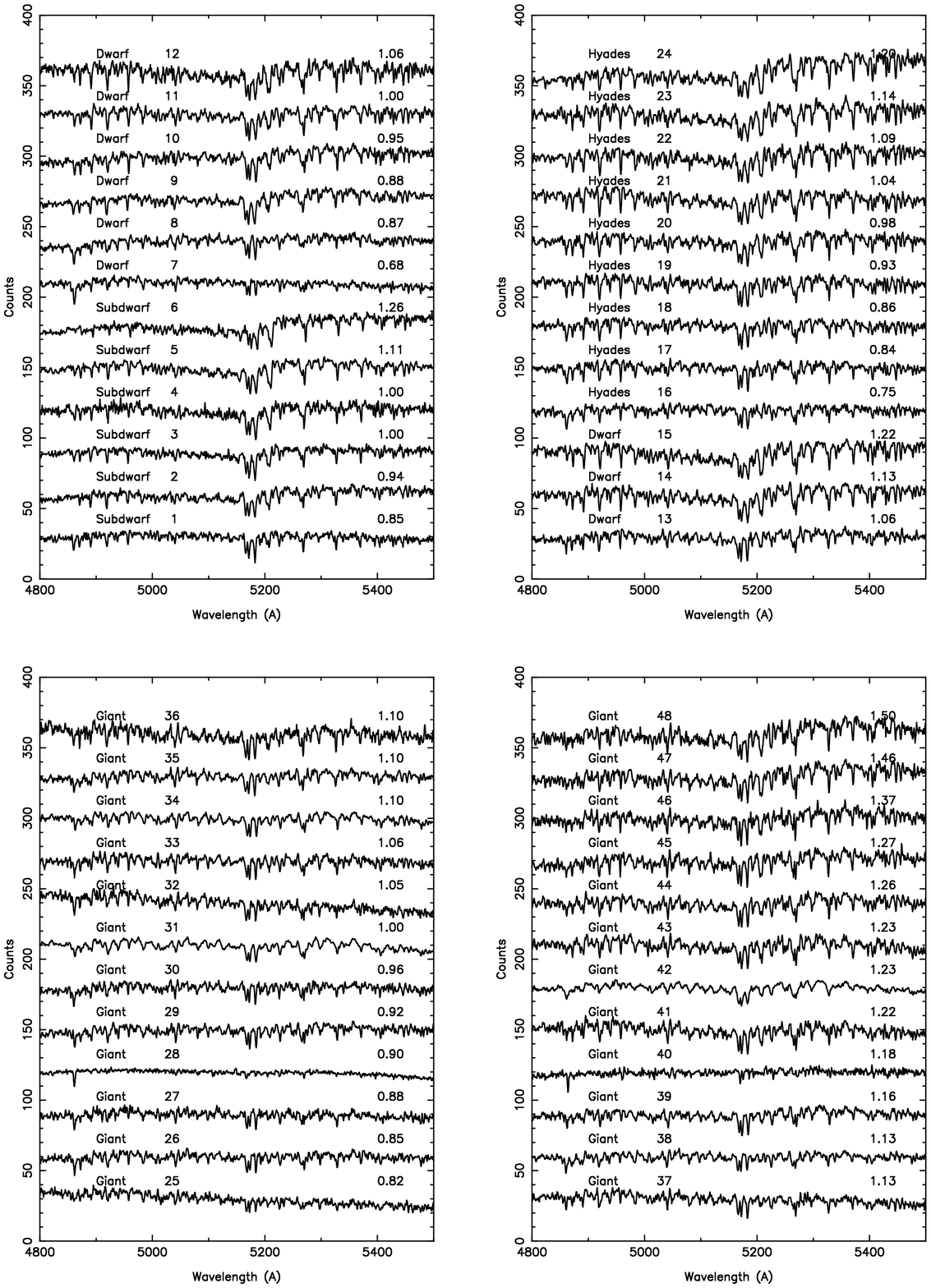}} 
\caption{ The grid of spectral standard stars.  Each star  is labeled with a
classification (subdwarf, dwarf, Hyades star, giant), with an identification
number (1 -- 48, corresponding to the entry in  Table~1), and with its ${\rm
(B-V)_0}$ color.}
\end{figure}

\begin{figure}
\epsfxsize=\hsize
\centerline{\epsffile{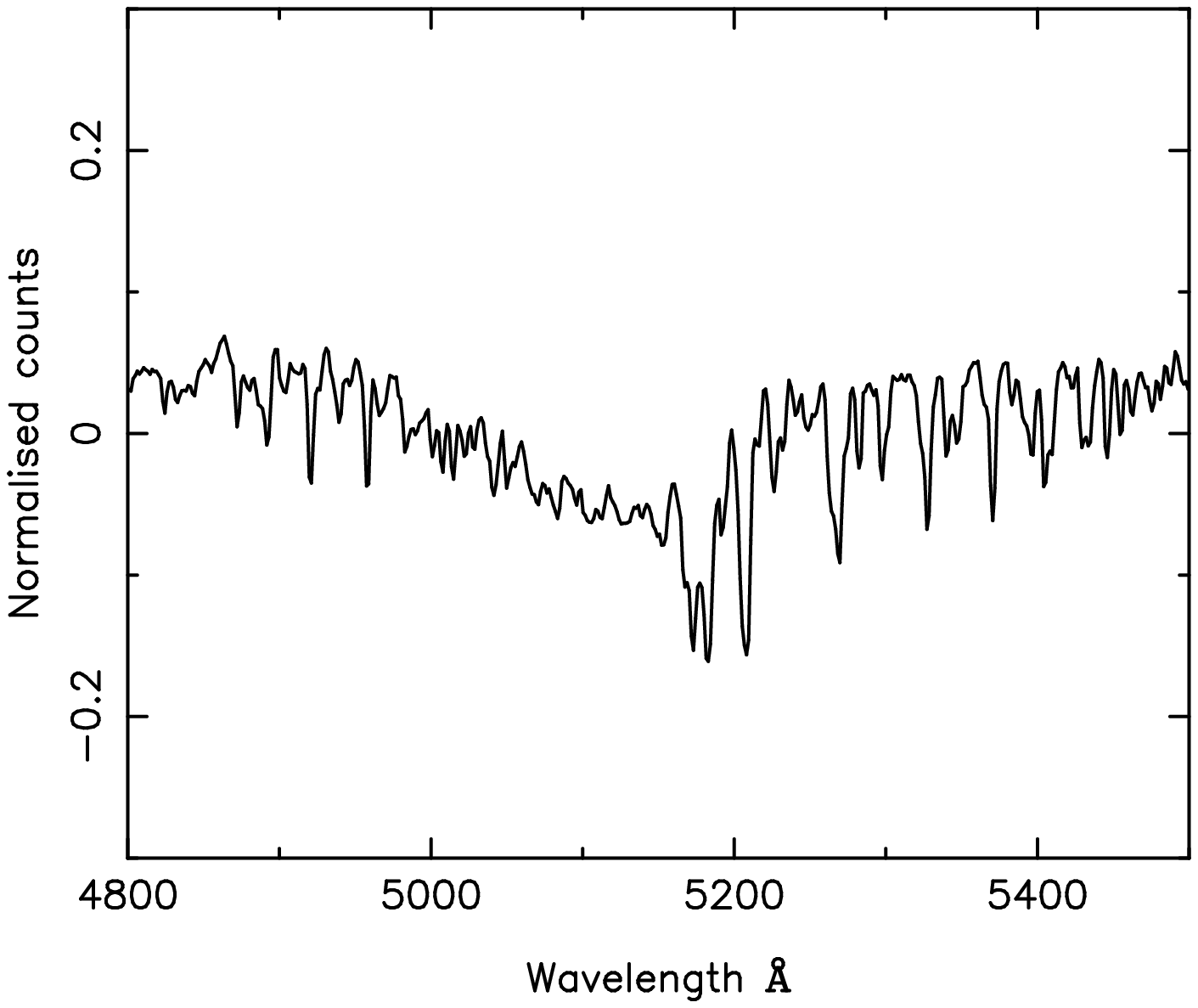}} 
\caption{The most discriminating  vector between K~dwarfs and K~giants (that
corresponding to $\lambda_1$ in Table~2).   The  dot product of this  vector
with a vector formed by the subtraction of a K~star spectrum and an averaged
spectrum of  standard stars gives a  color-dependent  parameter which can be
calibrated for  surface gravity.  Many  stellar features are visible in this
vector:  the  Mg`b'  feature  (a  wide  molecular  absorption  band that  is
strongest at   $\approx 5170 \AA$),  the  MgH band at $5205\AA$  and several
prominent Fe lines.}
\end{figure}

\vfill\eject

\begin{figure}
\epsfxsize=\hsize
\centerline{\epsffile{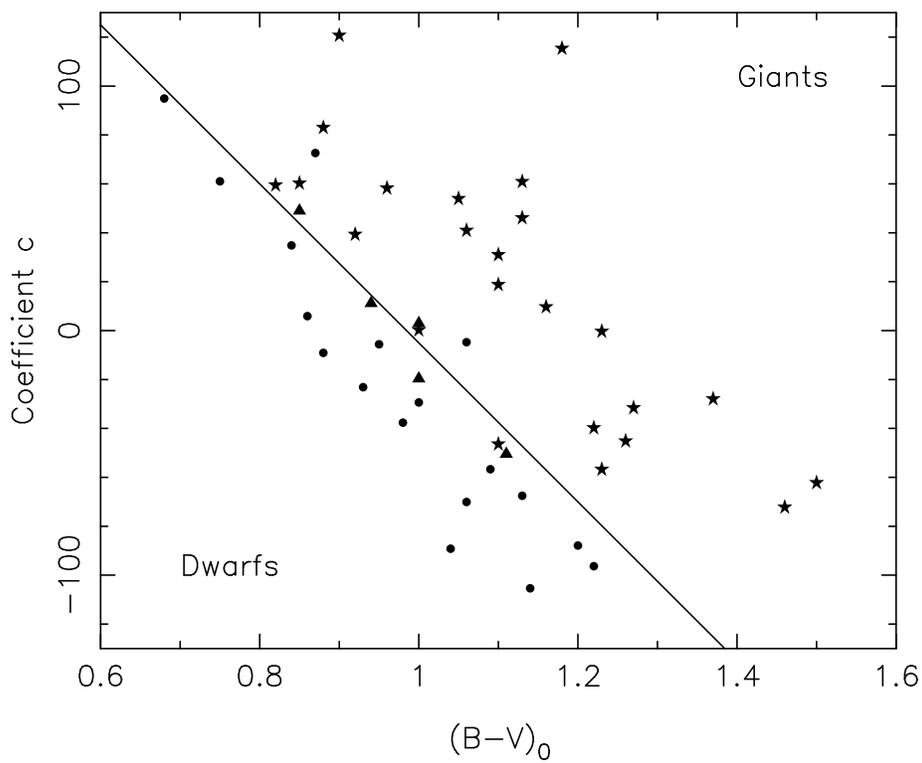}} 
\caption{The color  dependence of the coefficient $c$  (defined in the text)
with ${\rm  (B-V)_0}$ for the standards shown  in Figure~1.   The `stars' in
the diagram are  giants, the  `filled circles'  are dwarfs  and the  `filled
triangles' are subdwarfs.}
\end{figure}

\vfill\eject

\begin{figure}
\epsfxsize=\hsize
\centerline{\epsffile{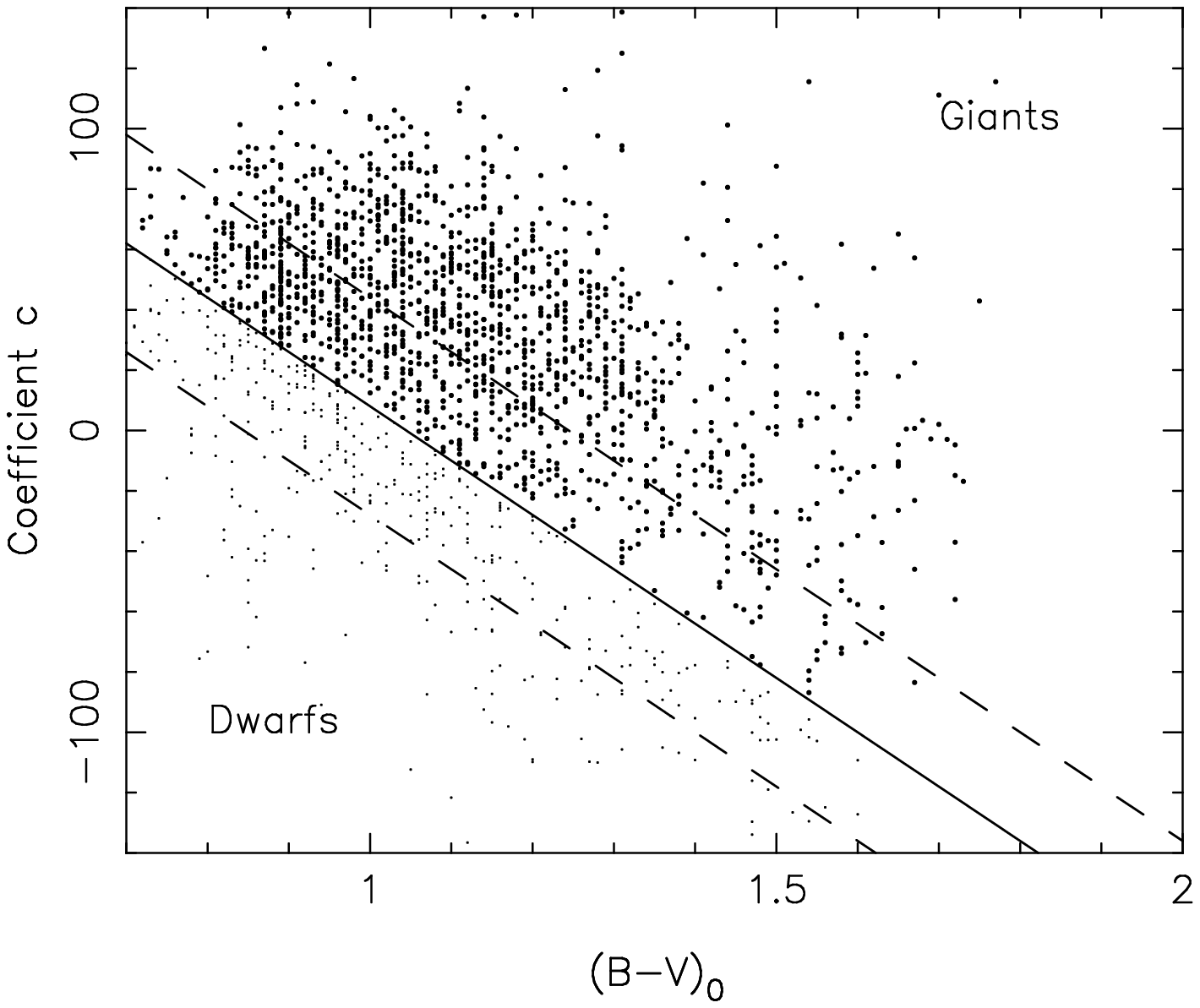}} 
\caption{Spectral classification  of survey   stars.  Given the  giant-dwarf
boundary shown in   Figure~3, the  survey stars  have  been classified  into
`large  dots'-giants and `dots'-dwarfs.   The values  of ${\rm (B-V)_0}$ are
from  APM  photometry (described in Ibata  \&  Gilmore 1995), with reddening
taken from Burstein  \& Heiles (1982) maps.   The dashed lines correspond to
$1\sigma$ photometric error deviations.}
\end{figure}

\vfill\eject
\begin{deluxetable}{ccccccccc}
\footnotesize
\tablecaption{The grid of standard stars}
\tablewidth{0pt}
\tablehead{
\colhead{No.} & \colhead{Class}   & \colhead{${\rm B-V}$}   & \colhead{star} & 
\colhead{No.} & \colhead{Class}   & \colhead{${\rm B-V}$}   & \colhead{star} &
\colhead{${\rm [Fe/H]}$}
}
\startdata
 1  &  subdwarf      &   0.85 &  LFT 1756 &
25  &  giant         &   0.82 &  HD 191179  &  \nl
 2  &  subdwarf      &   0.94 &  BD--$0^\circ$4234  &
26  &  giant         &   0.85 &  HD 201195  &  \nl
 3  &  subdwarf      &   1.00 &  LFT 466  &
27  &  giant         &   0.88 &  HD 203066  &  \nl
 4  &  subdwarf      &   1.00 &  G 155-35  &
28  &  giant         &   0.90 &  HR 5270    & -2.60  \nl
 5  &  subdwarf      &   1.11 &  LFT 100  &
29  &  giant         &   0.92 &  HD 171391  &  \nl
 6  &  subdwarf      &   1.26 &  LFT 1668  &
30  &  giant         &   0.96 &  HD 192246  &  \nl
 7  &  dwarf         &   0.68 &  HR 72  &
31  &  giant         &   1.00 &  HR 3994    &  ~0.22  \nl
 8  &  dwarf         &   0.87 &  HR 7703  &
32  &  giant         &   1.05 &  HD 202978  &  \nl
 9  &  dwarf         &   0.88 &  HR 487  &
33  &  giant         &   1.06 &  HD 157457  &  \nl
10  &  dwarf         &   0.95 &  HR+$21^\circ$3245  &
34  &  giant         &   1.10 &  HR 4287    &   -0.06  \nl
11  &  dwarf         &   1.00 &  HR 8382  &
35  &  giant         &   1.10 &  HR 8841    &   -0.13  \nl
12  &  dwarf         &   1.06 &  HR 8387  &
36  &  giant         &   1.10 &  HR 8924    &   ~0.55  \nl
13  &  dwarf         &   1.06 &  BD+$10^\circ$3665  &
38  &  giant         &   1.13 &  HR 7430    &   -0.70  \nl
14  &  dwarf         &   1.13 &  BD+$22^\circ$3406  &
37  &  giant         &   1.13 &  HD 211475  &  \nl
15  &  dwarf         &   1.22 &  BD+$6^\circ$4741  &
39  &  giant         &   1.16 &  HD 107328  &   -0.47  \nl
16  &  Hyades        &   0.75 &  vB 69  &
40  &  giant         &   1.18 &  HD 110184  &   -2.50  \nl
17  &  Hyades        &   0.84 &  Pels 50  &
41  &  giant         &   1.22 &  HD 202168  &  \nl
18  &  Hyades        &   0.86 &  Pels 56  &
42  &  giant         &   1.23 &  HR 5340    &   -0.42  \nl
19  &  Hyades        &   0.93 &  Pels 52  &
43  &  giant         &   1.23 &  HR 5370    &   ~0.31  \nl
20  &  Hyades        &   0.98 &  Pels 63  &
44  &  giant         &   1.26 &  HR 5582    &   ~0.42  \nl
21  &  Hyades        &   1.04 &  Pels 39  &
45  &  giant         &   1.27 &  HD 201875  &  \nl
22  &  Hyades        &   1.09 &  Pels 51  &
46  &  giant         &   1.37 &  HR 0489    &   -0.11  \nl
23  &  Hyades        &   1.14 &  Pels 49  &
47  &  giant         &   1.46 &  HR 6136    &   ~0.35  \nl
24  &  Hyades        &   1.20 &  Pels 65  &
48  &  giant         &   1.50 &  HR 0224    &   -0.07  \nl
\enddata
\end{deluxetable}

\vfill\eject
\begin{deluxetable}{cccc}
\footnotesize
\tablecaption{The ten largest eigenvalues of the covariance matrix of standard 
stars.}
\tablewidth{0pt}
\tablehead{
\colhead{No.} & \colhead{$\lambda$}  & \colhead{\% of trace}   & 
\colhead{cum. \% of trace}}
\startdata
~1  &  7027827.5  &  50.145     &  50.145 \nl
~2  &  1836905.6  &  13.107     &  63.252 \nl
~3  &  1277226.4  &  ~9.113     &  72.365 \nl
~4  &  ~606478.0  &  ~4.327     &  76.692 \nl
~5  &  ~295777.4  &  ~2.110     &  78.803 \nl
~6  &  ~251045.7  &  ~1.791     &  80.594 \nl
~7  &  ~207668.2  &  ~1.482     &  82.076 \nl
~8  &  ~159497.0  &  ~1.138     &  83.214 \nl
~9  &  ~149020.3  &  ~1.063     &  84.277 \nl
10  &  ~146001.9  &  ~1.042     &  85.319 \nl
\enddata
\end{deluxetable}

\end{document}